\documentclass[twocolumn,prl]{revtex4}
\usepackage{amsfonts}
\usepackage[T1]{fontenc}
\usepackage{amsmath,amsbsy,amssymb,graphicx}
\usepackage{times}
\usepackage{amsmath}
\usepackage{amssymb}
\usepackage{graphicx}

\let\mathbf=\boldsymbol

\begin{document}

\title{{\Large Topological Phase Transition and Electrically Tunable
Diamagnetism in Silicene}}
\author{Motohiko Ezawa}
\affiliation{Department of Applied Physics, University of Tokyo, Hongo 7-3-1, 113-8656,
Japan }

\begin{abstract}
Silicene is a monolayer of silicon atoms forming a honeycomb lattice. The
lattice is actually made of two sublattices with a tiny separation. Silicene
is a topological insulator, which is characterized by a full insulating gap
in the bulk and helical gapless edges. It undergoes a phase transition from
a topological insulator to a band insulator by applying external electric
field. Analyzing the spin Chern number based on the effective Dirac theory,
we find their origin to be a pseudospin meron in the momentum space. The
peudospin degree of freedom arises from the two-sublattice structure. Our
analysis makes clear the mechanism how a phase transition occurs from a
topological insulator to a band insulator under increasing electric field.
We propose a method to determine the critical electric field with the aid of
diamagnetism of silicene. Diamagnetism is tunable by the external electric
field, and exhibits a singular behaviour at the critical electric field. Our
result is important also from the viewpoint of cross correlation between
electric field and magnetism. Our finding will be important for future
electro-magnetic correlated devices.
\end{abstract}

\maketitle


\section{Introduction}

Silicene has recently been synthesized\cite{Lalmi,Padova,Aufray,GLayPRL} and
attracted much attention\cite{LiuPRL,EzawaNJP,EzawaJ}. It is a monolayer of
silicon atoms forming a two-dimensional honeycomb lattice. Almost every
striking property of graphene could be transferred to this innovative
material. Indeed, its low-energy dynamics is described by the Dirac theory
as in graphene. However, Dirac electrons are massive due to a relatively
large spin-orbit (SO) gap of $1.55$meV in silicene, where the mass can be
controlled by applying the electric field $E_{z}$ perpendicular to the
silicene sheet\cite{EzawaNJP}. A novel feature is that silicene is a
topological insulator\cite{LiuPRL}, which is characterized by a full
insulating gap in the bulk and helical gapless edges.

\begin{figure}[t]
\centerline{\includegraphics[width=0.4\textwidth]{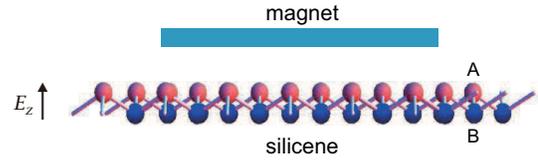}}
\caption{(Color online) Illustration of an experimental setting to determine
the critical electric field $E_{\text{cr}}$ with the use of the magnetic
susceptibility of silicene. Silicene consists of a honeycomb lattice of
silicon atoms with two sublattices made of A sites (red) and B sites (blue).
The two sublattice planes are separated by a distance. A strong diamagnetism
emerges between a magnet and a silicene sheet at the topological phase
transition point $E_{z}=E_{\text{cr}}$. }
\label{FigIllust}
\end{figure}

Silicene undergoes a topological phase transition from a topological
insulator to a band insulator as $|E_{z}|$ increases and crosses the
critical field $E_{\text{cr}}$, as has been shown\cite{EzawaNJP} by
examining numerically the emergence of the helical zero energy modes in
silicene nanoribbons. In this paper we present an analytic result by
calculating the topological numbers based on the effective Dirac theory. We
show that the origin of the topological numbers is a pseudospin meron in the
momentum space. The pseudospin degree of freedom arises from the
two-sublattice structure (Fig.\ref{FigIllust}). We also propose a simple
method to determine experimentally the phase transition point with the use
of the diamagnetism of silicene (Fig.\ref{FigIllust}).

The magnetism of conventional metal is composed of the Pauli paramagnetism
due to the spin magnetic moment and the Landau diamagnetism due to the
orbital motion of electrons. The free electron system exhibits paramagnetism
since the magnitude of the spin component is larger than the orbital
component. Contrarily, the Landau diamagnetism overcomes the Pauli
paramagnetism in a certain condensed matter system. An extreme case is
provided by graphene\cite{McClure,Screening,Koshino2010,Nakamura,Arimura},
where the orbital susceptibility has a strong singularity due to the gapless
character of Dirac electrons.

We calculate the magnetic susceptibility of silicene as a function of the
electric field $E_{z}$. We expect to have a strong singularity at the
critical electric field, $|E_{z}|=E_{\text{cr}}$, since Dirac electrons
become gapless at this point\cite{EzawaNJP}. Indeed, we show that it
diverges at the critical electric field at zero temperature ($T=0$), though
the divergence is round off at finite temperature. However, it is clearly
observable as long as $k_{\text{B}}T\lesssim \frac{1}{10}\lambda _{\text{SO}}
$. Our result is important also from the viewpoint of cross correlation
between electric field and magnetism. In general electric-field-controled
magnetism is rather difficult compared to magnetic-field-controled
electricity. On the other hand, the former is desirable since we can
precisely control electric field. Our finding will be important for future
electro-magnetic correlated devices.

\section{Tight Binding Model}

Silicene consists of a honeycomb lattice of silicon atoms with two
sublattices made of A sites and B sites. The two sublattices are separated
by a distance, which we denote by $2\ell $ with $\ell =0.23$\AA . The
silicene system is described by the four-band second-nearest-neighbor tight
binding model\cite{KaneMele,LiuPRB,EzawaNJP},

\begin{align}
H_{0}=& -t\sum_{\left\langle i,j\right\rangle \alpha }c_{i\alpha }^{\dagger
}c_{j\alpha }+i\frac{\lambda _{\text{SO}}}{3\sqrt{3}}\sum_{\left\langle
\!\left\langle i,j\right\rangle \!\right\rangle \alpha \beta }\nu
_{ij}c_{i\alpha }^{\dagger }\sigma _{\alpha \beta }^{z}c_{j\beta }  \notag \\
& -i\frac{2}{3}\lambda _{\text{R2}}\sum_{\left\langle \!\left\langle
i,j\right\rangle \!\right\rangle \alpha \beta }\mu _{i}c_{i\alpha }^{\dagger
}\left( \mathbf{\sigma }\times \hat{\mathbf{d}}_{ij}\right) _{\alpha \beta
}^{z}c_{j\beta },  \label{BasicHamil0}
\end{align}%
where $c_{i\alpha }^{\dagger }$ creates an electron with spin polarization $%
\alpha $ at site $i$, and $\left\langle i,j\right\rangle /\left\langle
\!\left\langle i,j\right\rangle \!\right\rangle $ run over all the
nearest/next-nearest neighbor hopping sites. The first term represents the
usual nearest-neighbor hopping with the transfer energy $t=1.6$eV. The
second term represents the effective SO coupling with $\lambda _{\text{SO}%
}=3.9$meV, where $\mathbf{\sigma }=(\sigma _{x},\sigma _{y},\sigma _{z})$ is
the Pauli matrix of spin, with $\nu _{ij}=+1$ if the
next-nearest-neighboring hopping is anticlockwise and $\nu _{ij}=-1$ if it
is clockwise with respect to the positive $z$ axis. The third term
represents the second Rashba SO coupling with $\lambda _{\text{R2}}=0.7$meV
associated with the next-nearest neighbor hopping term, where $\mu _{i}=\pm
1 $ for the A (B) site, and $\hat{\mathbf{d}}_{ij}=\mathbf{d}%
_{ij}/\left\vert \mathbf{d}_{ij}\right\vert $ with the vector $\mathbf{d}%
_{ij}$ connecting two sites $i$ and $j$ in the same sublattice.

We take a silicene sheet on the $xy$-plane, and apply the electric field $%
E_{z}$ perpendicular to the plane. There appear two additional terms in the
Hamiltonian,%
\begin{align}
H_{E}=& i\lambda _{\text{R1}}(E_{z})\sum_{\left\langle i,j\right\rangle
\alpha \beta }c_{i\alpha }^{\dagger }\left( \mathbf{\sigma }\times \hat{%
\mathbf{d}}_{ij}\right) _{\alpha \beta }^{z}c_{j\beta }  \notag \\
& +\ell \sum_{i\alpha }\mu _{i}E_{z}^{i}c_{i\alpha }^{\dagger }c_{i\alpha },
\label{BasicHamilE}
\end{align}%
where the first term represents the first Rashba SO coupling associated with
the nearest neighbor hopping, which is induced by external electric field%
\cite{Hongki,Tse}. It is proportional to the external electric field, $%
\lambda _{\text{R1}}(E_{z})\varpropto E_{z}$, and becomes of the order of $%
10\mu $eV at $E_{z}=\lambda _{\text{SO}}/\ell =17$meV\AA $^{-1}$. The second
term is the staggered sublattice potential term $\varpropto 2\ell E_{z}$
between silicon atoms at A sites and B sites. The total Hamiltonian is given
by $H=H_{0}+H_{E}$. We note that the first Rashba SO coupling term ($%
\varpropto \lambda _{\text{R1}}$) is missed in the previous analysis\cite%
{EzawaNJP}.

\begin{figure}[t]
\centerline{\includegraphics[width=0.27\textwidth]{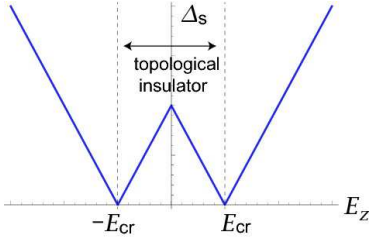}}
\caption{The band gap $\Delta _{s}$ as a function of the electric field $%
E_{z}$. The gap is open for $E_{z}\neq \pm E_{\text{cr}}$, where silicene is
an insulator. It has been shown\protect\cite{EzawaNJP} that it is a
topological insulator for $|E_{z}|<E_{\text{cr}}$ and a band insulator $%
|E_{z}|>E_{\text{cr}}$. Thus there occurs a topological phase transition at $%
|E_{z}|=E_{\text{cr}}$.}
\label{FigGap}
\end{figure}

\section{Dirac Theory}

Electronic states near the Fermi energy are $\pi $ orbitals residing near
the K and K' points at opposite corners of the hexagonal Brillouin zone. We
also call them the K$_{\eta }$ points with $\eta =\pm $. The low-energy
effective Hamiltonian is derived from the tight binding model $H=H_{0}+H_{E}$%
. It is described by the Dirac theory around the $K_{\eta }$ point as%
\begin{align}
H_{\eta }=& \hbar v_{\text{F}}\left( \eta k_{x}\tau _{x}+k_{y}\tau
_{y}\right) +\eta \tau _{z}h_{11}+\ell E_{z}\tau _{z}  \notag \\
& +\lambda _{\text{R1}}(\eta \tau _{x}\sigma _{y}-\tau _{y}\sigma _{x})/2
\label{DiracHamil}
\end{align}%
with 
\begin{equation}
h_{11}=\lambda _{\text{SO}}\sigma _{z}+a\lambda _{\text{R2}}\left(
k_{y}\sigma _{x}-k_{x}\sigma _{y}\right) ,
\end{equation}%
where $\tau _{a}$ is the Pauli matrix of the sublattice pseudospin, $v_{%
\text{F}}=\frac{\sqrt{3}}{2}at$ is the Fermi velocity, and $a=3.86$\r{A}\ is
the lattice constant.

The band gap is located at the K and K' points. At these points the energy
is exactly given by%
\begin{equation}
\mathcal{E}_{s}=\lambda _{\text{SO}}+s\ell E_{z},\quad \lambda _{\text{SO}}+s%
\sqrt{\left( \ell E_{z}\right) ^{2}+\lambda _{\text{R1}}^{2}}
\end{equation}%
where $s=\pm 1$ is the spin-chirality. It is given by $s=s_{z}\eta $ when
the spin $s_{z}$ is a good quantum number. The gap is given by $2|\Delta
_{s}\left( E_{z}\right) |$ with%
\begin{equation}
\Delta _{s}\left( E_{z}\right) =-s\lambda _{\text{SO}}+\frac{1}{2}\ell E_{z}+%
\frac{1}{2}\sqrt{\left( \ell E_{z}\right) ^{2}+\lambda _{\text{R1}}^{2}}.
\label{gapDiracX}
\end{equation}%
As $|E_{z}|$ increases, the gap decreases linearly since $\lambda _{\text{R1}%
}\propto E_{z}$, and vanishes at the critical point $|E_{z}|=E_{\text{cr}}$
with%
\begin{equation}
E_{\text{cr}}=\frac{s\lambda _{\text{SO}}}{\ell }\left[ 1-\left( \frac{%
\lambda _{\text{R1}}}{2\lambda _{\text{SO}}}\right) ^{2}\right] =\pm 17\text{%
meV/\AA },  \label{StepA}
\end{equation}%
and then increases linearly (Fig.\ref{FigGap}) The correction due to the
first Rashba coupling is extremely small, $(\lambda _{\text{R1}}/2\lambda _{%
\text{SO}})^{2}=10^{-4}$.

\begin{figure}[t]
\centerline{\includegraphics[width=0.5\textwidth]{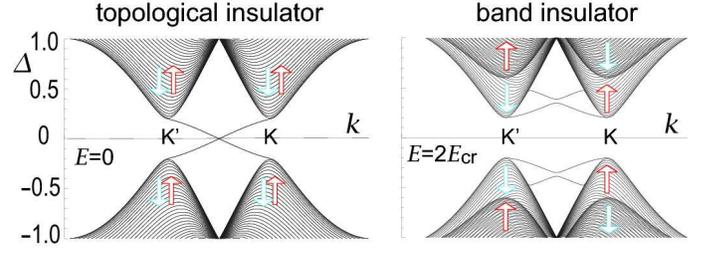}}
\caption{(Color online) One-dimensional energy bands for a silicene
nanoribbon. (a) The bands crossing the gap represent edge states,
demonstrating that it is a topological insulator. (b) All states are gapped,
demonstrating that it is a band insulator.}
\label{FigSRibbonBand}
\end{figure}

\section{Spin Chern Number}

We have shown in a previous paper\cite{EzawaNJP} that silicene is a
topological insulator for $|E_{z}|<E_{\text{cr}}$ and it is a band insulator
for $|E_{z}|<E_{\text{cr}}$ by examining numerically the emergence of the
helical zero energy modes in a silicene nanoribbon (Fig.\ref{FigSRibbonBand}%
). In this section we present an analytic discussion by calculating the
topological numbers based on the effective Dirac theory.

The topological quantum numbers are the Chern number $\mathcal{C}$ and the $%
\mathbb{Z}_{2}$ index. If the spin $s_{z}$ is a good quantum number, the $%
\mathbb{Z}_{2}$ index is identical to the spin-Chern number $\mathcal{C}_{s}$
modulo $2$. They are defined when the state is gapped, and given by 
\begin{equation}
\mathcal{C}=\mathcal{C}_{+}+\mathcal{C}_{-},\qquad \mathcal{C}_{s}=\frac{1}{2%
}(\mathcal{C}_{+}-\mathcal{C}_{-}),  \label{ChernB}
\end{equation}%
where $\mathcal{C}_{\pm }$ is the summation of the Berry curvature in the
momentum space over all occupied states of electrons with $s_{z}=\pm 1$.
They are well defined even if the spin is not a good quantum number\cite%
{Prodan09B,Yang}. In the present model the spin is not a good quantum number
because of spin mixing due to the Rashba couplings $\lambda _{\text{R1}}$
and $\lambda _{\text{R2}}$, and the resulting angular momentum eigenstates
are indexed by the spin chirality $s=\pm 1$. A convenient way of calculating
the Chern number $\mathcal{C}$ and the $\mathbb{Z}_{2}$ index is to use the
formula (\ref{ChernB}) to the system without the Rashba couplings and then
adiabatically switching on these couplings to recover the present system\cite%
{Prodan09B,Yang}.

When we set $\lambda _{\text{R1}}=0$ and $\lambda _{\text{R2}}=0$, the
Hamiltonian (\ref{DiracHamil}) becomes block diagonal. For each spin $%
s_{z}=\pm 1$ and valley $\eta =\pm 1$, it describes a two-band system in the
form,%
\begin{equation}
H=\mathbf{\tau }\cdot \mathbf{d},
\end{equation}%
where 
\begin{equation}
d_{x}=\eta \hbar v_{\text{F}}k_{x},\quad d_{y}=\hbar v_{\text{F}}k_{y},\quad
d_{z}=m_{\text{D}},
\end{equation}%
with the Dirac mass%
\begin{equation}
m_{\text{D}}=s\lambda _{\text{SO}}+\ell E_{z}.
\end{equation}%
The summation of the Berry curvature is reduced to the Pontryagin index in
the two-band system\cite{Qi},%
\begin{equation}
\mathcal{C}_{s_{z}}^{\eta }=\frac{1}{4\pi }\int d^{2}k\left( \frac{\partial 
\mathbf{\hat{d}}}{\partial k_{x}}\times \frac{\partial \mathbf{\hat{d}}}{%
\partial k_{y}}\right) \cdot \mathbf{\hat{d}}  \label{Pontryagin}
\end{equation}%
where $\mathbf{\hat{d}}=\mathbf{d/}\left\vert \mathbf{d}\right\vert $ is the
unit vector which specifies the direction of $\mathbf{d}$. It is equal to
the number of times the unit sphere is covered upon integrating over the
Brillouin zone.

\begin{figure}[t]
\centerline{\includegraphics[width=0.27\textwidth]{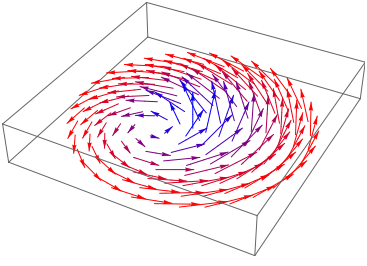}}
\caption{Illustration of a meron in momentum space. The pseudospin texture
is a meron for each spin in each valley, which yields $1/2$ to the
Pontryagin number.}
\label{meron}
\end{figure}

It is convenient to use the cylindrical coordinate in the momentum space,
where%
\begin{equation}
\hat{d}_{x}\pm i\hat{d}_{y}=\sqrt{1-\sigma ^{2}(k)}e^{i\eta \theta },\quad 
\hat{d}_{z}=\sigma (k)  \label{PpinText}
\end{equation}%
with%
\begin{equation}
\sigma (k)=\frac{m_{\text{D}}}{\sqrt{(\hbar v_{\text{F}}k)^{2}+m_{\text{D}%
}^{2}}},
\end{equation}%
The pseudospin texture (\ref{PpinText}) describes a vortex-like meron in the
momentum space, as shown in Fig.\ref{meron}. The Pontryagin index (\ref%
{Pontryagin}) yields 
\begin{equation}
\mathcal{C}_{s_{z}}^{\eta }={\frac{\eta }{4\pi }}\int \!d^{2}k\;\varepsilon
_{ij}\partial _{i}\sigma \partial _{j}\theta ={\frac{\eta }{2}}\text{sgn}%
\left( m_{\text{D}}\right) \int_{0}^{1}d\sigma .  \notag
\end{equation}%
Hence we find%
\begin{align}
\mathcal{C}=& \sum_{\eta =\pm }(\mathcal{C}_{+}^{\eta }+\mathcal{C}%
_{-}^{\eta })=0, \\
\mathcal{C}_{s}=& \sum_{\eta =\pm }\frac{1}{2}(\mathcal{C}_{+}^{\eta }-%
\mathcal{C}_{-}^{\eta })=\Theta (\lambda _{\text{SO}}-\ell |E_{z}|),
\end{align}%
where $\Theta $ is the step function, i.e., 
\begin{equation}
\mathcal{C}_{s}=\left\{ 
\begin{array}{cc}
1 & \text{for }\left\vert \ell E_{z}\right\vert <\lambda _{\text{SO}} \\ 
0 & \text{for }\left\vert \ell E_{z}\right\vert >\lambda _{\text{SO}}%
\end{array}%
\right. .
\end{equation}%
We have verified that the system is a topological insulator for $\left\vert
\ell E_{z}\right\vert <\lambda _{\text{SO}}$ and a band insulator for $%
\left\vert \ell E_{z}\right\vert >\lambda _{\text{SO}}$ in the system
without the Rashba interactions. The property remains true when they are
switched on adiabatically.

\section{Diamagnetism}

We proceed to discuss a possible experimental method to detect the phase
transition point by measuring the magnetic susceptibility. We apply
homogeneous magnetic field $\mathbf{B}=\mathbf{\nabla }\times \mathbf{A}%
=\left( 0,0,-B\right) $ with $B>0$ along the $z$ axis to silicene\cite%
{EzawaJ}. By making the minimal substitution, the Hamiltonian is given by%
\begin{align}
H_{\eta }=& v_{\text{F}}\left( \eta P_{x}\tau _{x}+P_{y}\tau _{y}\right)
+\eta \tau _{z}h_{11}+\ell E_{z}\tau _{z}  \notag \\
& +\lambda _{\text{R1}}(\eta \tau _{x}\sigma _{y}-\tau _{y}\sigma _{x})/2
\end{align}%
with the covariant momentum $P_{i}\equiv \hbar k_{i}+eA_{i}$. We introduce a
pair of Landau-level ladder operators, 
\begin{equation}
\hat{a}=\frac{\ell _{B}(P_{x}+iP_{y})}{\sqrt{2}\hbar },\quad \hat{a}%
^{\dagger }=\frac{\ell _{B}(P_{x}-iP_{y})}{\sqrt{2}\hbar },  \label{G-OperaA}
\end{equation}%
satisfying $[\hat{a},\hat{a}^{\dag }]=1$, where $\ell _{B}=\sqrt{\hbar /eB}$
is the magnetic length. In the basis $\left\{ \psi _{A\uparrow },\psi
_{B\uparrow },\psi _{A\downarrow },\psi _{B\downarrow }\right\} ^{t}$, the
Hamiltonian $H_{+}$ reads%
\begin{equation}
\left( 
\begin{array}{cccc}
\Delta _{+}^{0}\left( E_{z}\right) & \hbar \omega _{\text{c}}\hat{a}%
^{\dagger } & i\frac{\sqrt{2}\hbar a\lambda _{\text{R2}}}{\ell _{B}}\hat{a}%
^{\dagger } & 0 \\ 
\hbar \omega _{\text{c}}\hat{a} & -\Delta _{+}^{0}\left( E_{z}\right) & 
-i\lambda _{\text{R1}} & -i\frac{\sqrt{2}\hbar a\lambda _{\text{R2}}}{\ell
_{B}}\hat{a}^{\dagger } \\ 
-i\frac{\sqrt{2}\hbar a\lambda _{\text{R2}}}{\ell _{B}}\hat{a} & i\lambda _{%
\text{R1}} & \Delta _{-}^{0}\left( E_{z}\right) & \hbar \omega _{\text{c}}%
\hat{a}^{\dagger } \\ 
0 & i\frac{\sqrt{2}\hbar a\lambda _{\text{R2}}}{\ell _{B}}\hat{a} & \hbar
\omega _{\text{c}}\hat{a} & -\Delta _{-}^{0}\left( E_{z}\right)%
\end{array}%
\right) ,  \label{HamilBK}
\end{equation}%
at the K point, with $\omega _{\text{c}}=\sqrt{2}\hbar v_{\text{F}}/\ell
_{B} $. Here the diagonal elements $\Delta _{\pm }^{0}\left( E_{z}\right) $
are%
\begin{equation}
\Delta _{\pm }^{0}\left( E_{z}\right) =\pm \lambda _{\text{SO}}+\ell
E_{z}t_{z}
\end{equation}%
with the sublattice pseudospin $t_{z}$. We note that $\Delta _{\pm }\left(
E_{z}\right) =\Delta _{\pm }^{0}\left( E_{z}\right) $ if we set $\lambda _{%
\text{R1}}=0$.

\begin{figure}[t]
\centerline{\includegraphics[width=0.3\textwidth]{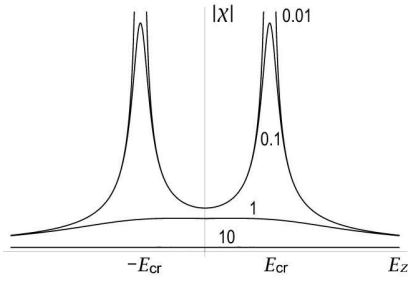}}
\caption{Susceptibility $\protect\chi $ as a function of electric field $%
E_{z}$ for various temperature $k_{\text{B}}T/\protect\lambda _{\text{SO}%
}=0.01,0.1,1,10$. It has a sharp peak at $|E_{z}|=E_{\text{cr}}$ for $k_{%
\text{B}}T/\protect\lambda _{\text{SO}}\lesssim 0.1$, and diverges as $%
T\rightarrow 0$.}
\label{FigSus}
\end{figure}

The magnetic susceptibility is defined by 
\begin{equation}
\chi =\lim_{B\rightarrow 0}\frac{M}{B},
\end{equation}%
where $M$ is the magnetization and $B$ is the external magnetic field. The
general formula for the orbital magnetic susceptibility of Bloch electrons
is given by\cite{Fukuyama},%
\begin{equation}
\chi =-g_{v}\frac{e^{2}\hbar ^{2}v_{\text{F}}^{2}}{6\beta \pi c^{2}}%
\sum_{n}\sum_{k}\text{Tr}\left( Gv_{x}Gv_{y}Gv_{x}Gv_{y}\right) ,
\label{Fukuyama}
\end{equation}%
where $v_{i}=\partial H/\partial k_{i}$, $\beta =1/k_{\text{B}}T$, and $G$
is the temperature Green function%
\begin{equation}
G\left( k,\omega _{n}\right) =\left( i\hbar \omega _{n}-H\right) ^{-1},
\end{equation}%
with $\hbar \omega _{n}=\left( 2n+1\right) \pi /\beta $ being the Matsubara
frequency.

By making the Taylor expansion of (\ref{Fukuyama}) with respect to $\lambda
_{\text{R1}}$ and $\lambda _{\text{R2}}$, the Rashba terms are found to
yield second order corrections, $\delta \chi =o(\lambda _{\text{R1}%
}^{2}/\hbar ^{2}v_{\text{F}}^{2})+o(\lambda _{\text{R2}}^{2}/\hbar ^{2}v_{%
\text{F}}^{2})$, and hence we neglect the Rashba terms in what follows. In
this approximation the spin $s_{z}$ is a good quantum number.

Integrating out the wave number $k$ of the matrix trace of (\ref{Fukuyama}
), we have 
\begin{equation}
\sum_{k}\text{Tr}\left( Gv_{x}Gv_{y}Gv_{x}Gv_{y}\right) =\sum_{s=\pm }\frac{%
2}{\hbar ^{2}\omega _{n}^{2}+\Delta _{s}^{2}}.
\end{equation}%
Using the formula of the infinite sum, 
\begin{equation}
\sum_{n}\frac{1}{\hbar ^{2}\omega _{n}^{2}+\Delta _{s}^{2}}=\frac{1}{2\Delta
_{s}k_{\text{B}}T}\tanh \frac{\Delta _{s}}{2k_{\text{B}}T},
\end{equation}%
we obtain the magnetic susceptibility at finite temperature,%
\begin{equation}
\chi \left( T,E_{z}\right) =-g_{v}\frac{e^{2}\hbar ^{2}v_{\text{F}}^{2}}{%
6\pi c^{2}}\sum_{s=\pm }\frac{1}{\Delta _{s}}\tanh \frac{\Delta _{s}}{k_{%
\text{B}}T}.  \label{FiniteSus}
\end{equation}%
where $g_{v}=2$ represents the valley degree of freedom. We show $\chi
\left( T,E_{z}\right) $ as a function of $E_{z}$ for typical values of $T$.

We can make a precise determination of the critical electric field $E_{\text{%
cr}}$ based on this formula. Let us consider, for example, a situation where
a magnet is placed parallel to a silicene sheet (Fig.\ref{FigIllust}). The
magnetization of silicene is given by $M=-|\chi |B$, where $B$ is the
external magnetic field made by the magnet. The magnet feels a strong
repulsive force at $E_{z}=E_{\text{cr}}$, since we have $\Delta
_{s}\rightarrow 0$ and $M\rightarrow -\infty $ as $|E_{z}|\rightarrow E_{%
\text{cr}}$ at $T=0$. Even for finite temperature, $\chi \left(
T,E_{z}\right) $ has a sharp peak at $|E_{z}|=E_{\text{cr}}$ for $k_{\text{B}%
}T\lesssim \lambda _{\text{SO}}/10$ as in Fig.\ref{FigSus}. Such a strong
repulsive force can be detected mechanically.

I am very much grateful to N. Nagaosa for many fruitful discussions on the
subject. This work was supported in part by Grants-in-Aid for Scientific
Research from the Ministry of Education, Science, Sports and Culture No.
22740196.

\end{document}